\documentclass[conference]{IEEEtran}

\IEEEoverridecommandlockouts
\usepackage{cite}
\usepackage{amsmath,amssymb,amsfonts}
\usepackage{graphicx}
\usepackage{ulem}
\usepackage{textcomp}
\usepackage{xcolor}
\usepackage{multirow}
\usepackage{longtable}
\usepackage{subcaption,lipsum}
\usepackage[flushleft]{threeparttable}
\usepackage{adjustbox}
\usepackage{graphicx}
\usepackage{tikz}
\usepackage{array}
\usepackage{listings}
\usepackage{soul}
\usepackage[most]{tcolorbox}
\usepackage{balance}
\usepackage{enumitem}
\usepackage{xurl}
\usepackage{hyperref}
\usepackage{xcolor}
\usepackage{algorithm}
\usepackage{algpseudocode}
\usepackage{stmaryrd}
\usepackage{color, colortbl}
\usepackage{amsfonts} 
\usepackage{booktabs}
\usepackage{caption}
\usepackage{epigraph} 
\usepackage{framed}
\usepackage{here}
\usepackage{ulem}
\usepackage{makecell}
\usepackage{csquotes}
\usepackage{afterpage}
\usepackage{datenumber}

\definecolor{chestnut}{rgb}{0.8, 0.36, 0.36}

\definecolor{chestnut}{rgb}{0.8, 0.36, 0.36}

\newcounter{dateone}\newcounter{datetwo}%
\newcommand{\daydifftoday}[3]{%
\setmydatenumber{dateone}{\the\year}{\the\month}{\the\day}%
\setmydatenumber{datetwo}{#1}{#2}{#3}%
\addtocounter{datetwo}{-\thedateone}%
\thedatetwo
}
\definecolor{LHScolor}{HTML}{555555}
\definecolor{ABlue}{HTML}{127bca}
\newcommand{\droptextshadow}[2]{%
    \tikz[baseline,outer sep=0pt, inner sep=0pt]{
    \node[#1!40!black] at (0,-0.1ex) {#2};
    \node[white] at (0,0) {#2};
}%
}
\newcommand{\DOIbox}[1]{
\tcbsidebyside[
        bicolor,
        sidebyside,
        sidebyside adapt=both,
        sidebyside gap=5pt,
        top=0pt,left=0pt,right=0pt,bottom=0pt,
        boxrule=0pt,rounded corners,
        interior style={top color=LHScolor,bottom color=LHScolor!60!black},
        segmentation style={top color=ABlue,bottom color=ABlue!60!black},
]{%
\droptextshadow{LHScolor}{DOI}% <-- Drop shadow + text for DOI 
}{%
\droptextshadow{ABlue}{\href{http://dx.doi.org/#1}{#1}}% <-- Drop shadow + text for reference number + hyperref
}%
}

\newcommand{\RqOne}{\textbf{RQ1:} \emph{What are differences between documentation and code for human-written notebook?}}

\newcommand{\RqTwo}{\textbf{RQ2:} \emph{What are the key differences between GenAI and human-written data science notebooks?}}

\definecolor{Large}{HTML}{696969}
\definecolor{Negligible}{HTML}{D3D3D3}
\definecolor{Medium}{HTML}{808080}
\definecolor{Small}{HTML}{A9A9A9}
\definecolor{background}{HTML}{F7F7F7} % light gray background for code
\definecolor{commentcolor}{HTML}{6A737D} % gray comment color for commit message
\definecolor{green}{HTML}{28A745} % green for add
\definecolor{red}{HTML}{D73A49} % red for delete
\definecolor{blue}{HTML}{0366D6} % blue for branch names, file names, etc.
\definecolor{coolblack}{rgb}{0.0, 0.18, 0.39}
\definecolor{awesome}{rgb}{0.0, 0.2, 0.6}

\def\BibTeX{{\rm B\kern-.05em{\sc i\kern-.025em b}\kern-.08em
    T\kern-.1667em\lower.7ex\hbox{E}\kern-.125emX}}
\begin{document}

\title{Human to Document, AI to Code: \\ Comparing GenAI for Notebook Competitions}

\author{
    \IEEEauthorblockN{
        Tasha Settewong\IEEEauthorrefmark{1}, 
         Youmei Fan\IEEEauthorrefmark{1},
         Raula Gaikovina Kula\IEEEauthorrefmark{2},
         Kenichi Matsumoto\IEEEauthorrefmark{1}
    }
    \IEEEauthorblockA{
        \IEEEauthorrefmark{1}Nara Institute of Science and Technology, Japan\\
        \{tasha.settewong.ts1, fan.youmei.fs2, matumoto\}@is.naist.jp
    }
    \IEEEauthorblockA{
        \IEEEauthorrefmark{2}The University of Osaka, Japan\\
        raula-k@ist.osaka-u.ac.jp
    }
}

\maketitle

\begin{abstract}
 Computational notebooks have become the preferred tool of choice for data scientists and practitioners to perform analyses and share results.
 Notebooks uniquely combine scripts with documentation.
 With the emergence of generative AI (GenAI) technologies, it is increasingly important, especially in competitive settings, to distinguish the characteristics of human-written versus GenAI.
 Our new idea is to explore the strengths of both humans and GenAI through the coding and documenting activities in notebooks. We first characterize differences between 25 code and documentation features in human-written, medal-winning Kaggle notebooks. We find that gold medalists are primarily distinguished by longer and more detailed documentation. Second, we analyze the distinctions between human-written and GenAI notebooks. Our results show that while GenAI notebooks tend to achieve higher code quality (as measured by metrics like code smells and technical debt), human-written notebooks display greater structural diversity, complexity, and innovative approaches to problem-solving. 
 Based on these early results, we highlight four agendas to further investigate how GenAI could be utilized in notebooks that maximize the potential collaboration between human and GenAI tech.

\end{abstract}


\section{Introduction}
Computational notebooks have rapidly become an important tool for data scientists and researchers. According to Jupyter, a notebook is a shareable document that combines code, text, data, and rich visualizations, offering an interactive environment for prototyping, data exploration, and sharing ideas, so that users can learn coding and documentation\footnote{\url{https://docs.jupyter.org/en/latest/}}. Platforms like Kaggle have further accelerated the adoption of notebooks, offering a space for machine learning competitions that attract both industry practitioners and academic researchers \cite{perkel2018jupyter}. As Kaggle\footnote{\url{https://www.kaggle.com/}} competitions become more popular and prestigious, participants face increasingly tougher competition to create the best, most insightful notebooks.

The landscape of coding has been transformed by the emergence of generative AI (GenAI) technologies such as ChatGPT\footnote{\url{https://chatgpt.com/}}, Gemini\footnote{\url{https://gemini.google.com/}} Claude\footnote{\url{https://claude.ai/}} and open models from Meta Ollama\footnote{\url{https://ollama.com/}} and Huggingface\footnote{\url{https://huggingface.co/}}. 
These tools have the potential to assist practitioners by generating code, writing explanations, and even producing entire notebooks with minimal human intervention. However, this technological shift raises important questions: How can we distinguish between notebooks created by humans and those generated by GenAI? More importantly, what can we learn from each to advance the state of the art in computational notebooks?
While prior work has examined code quality \cite{grotov2022large}, reproducibility \cite{pimentel2021understanding}, and collaboration \cite{wang2019data} in notebooks, the specific contributions and limitations of GenAI notebooks remain underexplored.
Competitions provide an ideal experimental setting, as participants are incentivized to submit their highest-quality work for rigorous community assessment and ranking. 

To investigate this phenomena, in this paper, we perform experiments on three case studies to compare human-written notebooks and those generated by leading large language models (LLMs).
The case studies is across three major Kaggle competitions (i.e.,  1. Santander Customer Transaction Prediction, 2. Home Credit Default Risk, and 3. IEEE-CIS Fraud Detection).
To differentiate between human-written notebooks, we use the heuristic of gold medals (i.e., notebooks voted as being high quality by the community) to identify quality notebooks\footnote{\url{https://www.kaggle.com/progression/}}. 
By extracting and analyzing 25 features related to documentation and code quality from a curated dataset of 465 human-authored notebooks and 9 GenAI notebooks, we answer two research questions: 
\begin{figure*}[t]
    \includegraphics[width=1.0\textwidth]{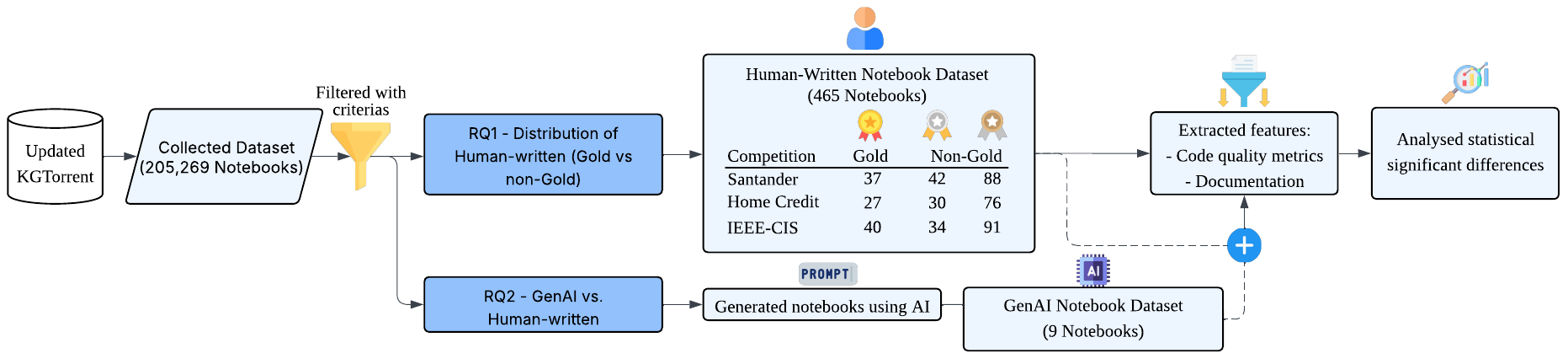}
    \centering
    \caption{Overview of Data Collection}
    \label{fig:overview}
\end{figure*}
\\
\RqOne\\
~The motivation behind RQ1 is to characterize the fundamental code and documentation patterns in human-written notebooks, revealing essential programming behaviors, documentation styles, and structural approaches that differentiate gold-medal human notebooks from the rest, establishing a crucial baseline.
\\
\RqTwo\\
~The motivation behind RQ2 is to identify the distinguishing factors that separate human and GenAI approaches in competitive data science, focusing on the impact of code quality metrics versus documentation on medal-worthy outcomes.

Our findings show that GenAI notebooks achieve higher code quality, with significantly fewer code smells, technical debt, and coding violations. However, human-written notebooks stand out for their more comprehensive and accessible documentation. In particular, gold medal-winning notebooks feature more than twice as much markdown content and narrative explanation as the non-gold group. Moreover, while GenAI documentation tends to require a higher reading level, human-written explanations are generally clearer and more approachable.

\section{Three Case Studies of Notebook Competitions}
Three closed competitions were selected for this study: Santander Customer Transaction Prediction\footnote{\url{https://www.kaggle.com/competitions/santander-customer-transaction-prediction}}, Home Credit Default Risk\footnote{\url{https://www.kaggle.com/competitions/home-credit-default-risk}}, and IEEE-CIS Fraud Detection\footnote{\url{https://www.kaggle.com/competitions/ieee-fraud-detection}}. 
The Santander Customer Transaction Prediction competition contributed 167 medal-winning notebooks to the dataset used in this analysis. From the Home Credit Default Risk competition, 133 notebooks were analyzed. The IEEE-CIS Fraud Detection competition provided the final 165 notebooks for the study.

To utilize the KGTorrent dataset effectively for GenAI notebook generation, we conducted a comprehensive investigation of the data structure. This analysis revealed that GenAI models require well-defined problem contexts to generate meaningful notebooks, necessitating the selection of specific Kaggle competitions that provide clear problem statements and structured datasets. 

The selection of these three competitions was based on the criterion that those with the highest number of participating teams were prioritized to maximize the sample size and ensure a robust statistical analysis.

\section{Study Design}

Figure~\ref{fig:overview} presents the workflow of the data collection process. We utilized the KGTorrent\cite{quaranta2021kgtorrent} tool to retrieve Jupyter notebooks. KGTorrent constitutes a comprehensive dataset encompassing computational notebooks (code kernels) and their associated metadata sourced from the Kaggle platform. 

To obtain an updated version of the KGTorrent dataset, Kaggle metadata\footnote{\url{https://www.kaggle.com/datasets/kaggle/meta-kaggle}} was downloaded on May 25th, 2025, and Jupyter notebooks were subsequently retrieved through API requests using the KGTorrent tool. This process yielded a total of 205,269 notebooks. The data collection process consists of four main stages: first, we identify target competitions for analysis; second, we apply data filtering criteria to select relevant notebooks; third, we generate corresponding GenAI notebooks using GenAI models; and finally, we extract features from both human-written and GenAI notebooks for comparative analysis.

\subsection{Data Collection and Filtering}
The filtering process yielded a total of 465 notebooks, with the distribution of medal achievements across the three competitions presented in Table~\ref{tab:dist-medal}. In this study, the notebook medals (i.e., Bronze, Silver, Gold) serve as proxies for computational notebook quality, as medal progression and achievement represent community recognition and appreciation of the work's merit.
Following competition selection, a systematic filtering process was implemented to ensure dataset integrity and completeness using the following criteria:
\begin{enumerate}
\item Both the notebook's metadata and corresponding notebook files must be present.
\item Contributor information for each notebook creator must be available.
\item The notebook must have been submitted to one of the three selected competitions.
\item The notebook must have received a medal recognition.
\end{enumerate}
\begin{table}[t]
\centering
\caption{Medal Distribution Across Competitions}
\label{tab:dist-medal}
\small
\renewcommand{\arraystretch}{1.3}
\begin{tabular}{lccc}
\toprule
\textbf{Medal Type} & \textbf{Home Credit} & \textbf{Santander} & \textbf{IEEE-CIS} \\
\midrule
Bronze & 76 & 88 & 91 \\
Silver & 30 & 42 & 34 \\
Gold & 27 & 37 & 40 \\
\midrule
\textbf{Total} & \textbf{133} & \textbf{167} & \textbf{165} \\
\bottomrule
\end{tabular}
\end{table}

% Define colors for effect size highlighting
\definecolor{negligible}{RGB}{240,248,255}
\definecolor{small}{RGB}{173,216,230}
\definecolor{medium}{RGB}{100,149,237}
\definecolor{large}{RGB}{65,105,225}

% \vspace{0.5em} % Small space before table

\definecolor{negligible}{RGB}{240,248,255}
\definecolor{small}{RGB}{173,216,230}
\definecolor{medium}{RGB}{100,149,237}
\definecolor{large}{RGB}{65,105,225}

\begin{table*}[t] % Force position to prevent floating
\centering
\caption{Statistical Comparisons of Gold and Non-Gold Human-Written Notebooks}
\label{tab-gold-nongold-compact}
\footnotesize
\renewcommand{\arraystretch}{1.55}
\setlength{\tabcolsep}{14pt}
\begin{tabular}{@{}l|cc|cc|cc@{}}
\hline
\multirow{2}{*}{\textbf{Feature}} & \multicolumn{2}{c|}{\textbf{Home-Credit}} & \multicolumn{2}{c|}{\textbf{Santander}} & \multicolumn{2}{c}{\textbf{IEEE-CIS}} \\
\cline{2-3} \cline{4-5} \cline{6-7}
& \textbf{p-value} & \boldmath$\eta^{2}$ & \textbf{p-value} & \boldmath$\eta^{2}$ & \textbf{p-value} & \boldmath$\eta^{2}$ \\
\hline
\#Markdown Char & \textbf{0.0005***} & \cellcolor{medium}0.085 & \textbf{0.0004***} & \cellcolor{medium}0.070 & -- & -- \\
\#Markdown Line & \textbf{0.001**} & \cellcolor{medium}0.074 & \textbf{0.0004***} & \cellcolor{medium}0.071 & -- & -- \\
\#Sentences & \textbf{0.001**} & \cellcolor{medium}0.073 & \textbf{0.0004***} & \cellcolor{medium}0.071 & -- & -- \\
\#Markdown Cell & \textbf{0.001**} & \cellcolor{medium}0.070 & \textbf{0.001***} & \cellcolor{small}0.056 & -- & -- \\
Avg Sentence & \textbf{0.013*} & \cellcolor{small}0.040 & -- & -- & -- & -- \\
Duplicated Lines & \textbf{0.021*} & \cellcolor{small}0.033 & \textbf{0.020*} & \cellcolor{small}0.026 & \textbf{0.004**} & \cellcolor{small}0.046 \\
Duplicated Blocks & -- & -- & \textbf{0.009**} & \cellcolor{small}0.035 & \textbf{0.004**} & \cellcolor{small}0.045 \\
\#Code Char & -- & -- & -- & -- & \textbf{0.011*} & \cellcolor{small}0.033 \\
\#Visual & -- & -- & -- & -- & \textbf{0.019*} & \cellcolor{small}0.027 \\
\hline
\end{tabular} % Controlled space before footnote
\begin{minipage}{\linewidth}
\vspace{1em}
\footnotesize
* p-value $< 0.05$; ** p-value $< 0.01$; and *** p-value $< 0.001$. The effect sizes with thresholds are highlighted in \colorbox{negligible}{Negligible} \colorbox{small}{Small} \colorbox{medium}{Medium} \colorbox{large}{Large}.
\end{minipage}
\end{table*}

% \vspace{-1.5em}
% Define colors for alternating rows
\definecolor{lightblue}{RGB}{240,248,255}
\definecolor{mediumblue}{RGB}{173,216,230}

\begin{table*}[t]
\centering
\caption{Gold vs Non-Gold Feature Statistics Comparisons}
\label{tab:gold-nongold-stats}
\footnotesize
% \small
\renewcommand{\arraystretch}{1.6}
\setlength{\tabcolsep}{14pt}
\begin{tabular}{@{}l|l|c|c|c|c|c@{}}
\hline
\textbf{Feature} & \textbf{Medal Type} & \textbf{Mean} & \textbf{Median} & \textbf{Std} & \textbf{Min} & \textbf{Max} \\
\hline
\multirow{2}{*}{\#Markdown Char} & Non-Gold & 2688.10 & 934.0 & 6526.72 & 0 & 94377 \\
 & \cellcolor{lightblue}Gold & \cellcolor{lightblue}5369.35 & \cellcolor{lightblue}2261.0 & \cellcolor{lightblue}8071.63 & \cellcolor{lightblue}0 & \cellcolor{lightblue}45281 \\
\hline
\multirow{2}{*}{\#Markdown Line} & Non-Gold & 35.02 & 16.0 & 67.15 & 0 & 748 \\
 & \cellcolor{lightblue}Gold & \cellcolor{lightblue}64.98 & \cellcolor{lightblue}35.5 & \cellcolor{lightblue}79.04 & \cellcolor{lightblue}0 & \cellcolor{lightblue}364 \\
\hline
\multirow{2}{*}{\#Sentences} & Non-Gold & 27.39 & 12.0 & 44.14 & 0 & 308 \\
 & \cellcolor{lightblue}Gold & \cellcolor{lightblue}56.46 & \cellcolor{lightblue}26.5 & \cellcolor{lightblue}75.81 & \cellcolor{lightblue}0 & \cellcolor{lightblue}364 \\
\hline
\multirow{2}{*}{\#Markdown Cell} & Non-Gold & 11.71 & 6.0 & 18.57 & 0 & 184 \\
 & \cellcolor{lightblue}Gold & \cellcolor{lightblue}21.12 & \cellcolor{lightblue}12.0 & \cellcolor{lightblue}27.64 & \cellcolor{lightblue}0 & \cellcolor{lightblue}160 \\
\hline
\end{tabular}
\end{table*}
% \vspace*{-1.5em}

\subsection{Generating notebook using GenAI}
To facilitate comparison between human-written and GenAI notebooks, three large language models (LLMs) were selected: (1) GPT-4.1, (2) Llama 4 Maverick, and (3) Gemini 2.5 Pro Review. These models were chosen from different companies (OpenAI, Meta, and Google ) to minimize potential bias and ensure diverse representation. A single, non-iterative prompt was used to establish a controlled baseline for the models' raw generative capabilities. This approach avoids introducing prompt engineering as a variable, which could create bias as different GenAI respond optimally to different prompting strategies.
\begin{quote}
\textit{``generate .ipynb file in JSON format for Kaggle competition the overview is \{comepetition\_overview\}.
\{evaluation\_method\}. Using these csv files; \{competition\_dataset\}"
\vspace{0.5em}
\begin{flushright}
-- \textit{Used prompt}
\end{flushright}}
\end{quote}
After .ipynb files in JSON format were generated by the GenAI models, the JSON format files were converted to standard Jupyter Notebook format and manually checked for file integrity and structural validity to ensure proper formatting. As part of a post-review validation to assess practical runnability, a basic execution is performed to check on all nine GenAI notebooks.

\subsection{Extracting features}
Individual features were extracted, resulting in a total of 25 features:

\textbf{(1) Documentation-related attributes} (\#Markdown Char, \#Markdown Line, \#Markdown Cell, \#Sentence, Avg Sentence, Gunning Fog, \#Visual, and Comment Lines*). 

\textbf{(2) Code quality-related attributes} (\#Code Char, \#Code Line, \#Code Cell, Cyclomatic Complexity*, Cognitive Complexity*, Functions*, Statements*, Duplicated Blocks*, Duplicated Lines*, Bugs*, Violations*, Code Smells*, Technical Debts*, Maintainability Rating*, Reliability Rating*, and Security Rating*)

This extraction process utilized the SonarQube\footnote{\url{https://sonarcloud.io/explore/projects}} tool, an open-source static analysis platform designed for continuous code quality monitoring that integrates analysis and reporting throughout the software development process \cite{9794341, campbell2013sonarqube, lenarduzzi2020survey} and is used to extract some features(*) through API.

\section{Empirical Study}
We now discuss the analysis method and results of the study.
\subsection{Answering RQ1}
\label{RQ1Methods}

To address RQ1, we assessed data normality through the Shapiro-Wilk test \cite{shapiro1965analysis} and analyzed statistically significant differences between notebook features and medal achievement using the Kruskal-Wallis H test \cite{kruskal1952use} with effect size calculations using the human-written notebooks dataset illustrated in Figure ~\ref{fig:overview}.

Table~\ref{tab-gold-nongold-compact} shows that gold medalists primarily differentiate themselves through documentation practices rather than coding metrics. Documentation features demonstrated the strongest discriminating power between gold and non-gold medalists, with markdown characters, lines, and sentence counts showing highly significant differences (p-value $\leq$ 0.001) and medium effect sizes (0.056-0.085). 

The distribution analyses in Table~\ref{tab:gold-nongold-stats} confirm that gold medalists consistently produce substantially longer documentation than non-gold medalists.
In contrast, code-related metrics showed weaker differentiation patterns. The IEEE-CIS competition displayed significance only in specific elements like duplicated lines and code character quantity, but with small effect sizes (0.033-0.046), indicating limited practical impact on distinguishing performance levels.
These findings indicate that documentation, rather than code quality, primarily separates gold medalists from other competitors. Gold medalists systematically generate notebooks with more extensive markdown content.

\begin{tcolorbox}[colback=gray!5,colframe=awesome,title= RQ1 Summary]
Higher quality human notebooks tend to have longer documentation. 
Gold medal worthy notebooks are distinguishable by having longer documentation (i.e., markdown characters, lines, and sentences).
\end{tcolorbox}

\subsection{Answering RQ2}
\label{subsection:RQ2}
% \input{tables/ai-gold}
% \input{tables/ai-nongold}
% Required packages:
% \usepackage{booktabs}
% \usepackage[table]{xcolor}
% \usepackage{graphicx}
% \usepackage{multirow}

% Define colors for effect sizes
% Define color scheme
\definecolor{negligible}{RGB}{240,248,255}
\definecolor{small}{RGB}{173,216,230}
\definecolor{medium}{RGB}{100,149,237}
\definecolor{large}{RGB}{65,105,225}

\begin{table*}[htbp]
\centering
\caption{Statistical Comparisons of GenAI and Each Medal Notebooks}
\label{tab:ai-each}
% \footnotesize
\small
\renewcommand{\arraystretch}{1.35}
\setlength{\tabcolsep}{10pt}
\begin{tabular}{l|l|cc|cc|cc}
\hline
& & \multicolumn{2}{c|}{\textbf{Home-Credit}} & \multicolumn{2}{c|}{\textbf{Santander}} & \multicolumn{2}{c}{\textbf{IEEE-CIS}} \\ 
\cline{3-4} \cline{5-6} \cline{7-8}
\textbf{Feature} & \textbf{Medal} & \textbf{p-value} & $\boldsymbol{\eta^{2}}$ & \textbf{p-value} & $\boldsymbol{\eta^{2}}$ & \textbf{p-value} & $\boldsymbol{\eta^{2}}$ \\ 
\hline
% Functions section
\multirow{3}{*}{Functions}
& Gold & \textbf{0.011*} & \cellcolor{large}0.197 & -- & -- & \textbf{0.041*} & \cellcolor{medium}0.078 \\
& Silver & \textbf{0.009**} & \cellcolor{large}0.191 & -- & -- & -- & -- \\
& Bronze & \textbf{0.023*} & \cellcolor{small}0.054 & -- & -- & -- & -- \\
\hline
% Statements section
\multirow{3}{*}{Statements}
& Gold & \textbf{0.025*} & \cellcolor{large}0.143 & -- & -- & -- & -- \\
& Silver & \textbf{0.018*} & \cellcolor{large}0.148 & -- & -- & -- & -- \\
& Bronze & -- & -- & -- & -- & -- & -- \\
\hline
% Comment Lines section
\multirow{3}{*}{Comment Lines}
& Gold & \textbf{0.035*} & \cellcolor{medium}0.122 & -- & -- & -- & -- \\
& Silver & -- & -- & -- & -- & -- & -- \\
& Bronze & -- & -- & -- & -- & -- & -- \\
\hline
% Code Cell section
\multirow{3}{*}{\#Code Cell}
& Gold & \textbf{0.049*} & \cellcolor{medium}0.103 & -- & -- & \textbf{0.025*} & \cellcolor{medium}0.098 \\
& Silver & -- & -- & -- & -- & -- & -- \\
& Bronze & -- & -- & -- & -- & -- & -- \\
\hline
% Cyclomatic Complexity section
\multirow{3}{*}{Cyclomatic Complexity}
& Gold & \textbf{0.050*} & \cellcolor{medium}0.102 & -- & -- & -- & -- \\
& Silver & \textbf{0.028*} & \cellcolor{medium}0.124 & -- & -- & -- & -- \\
& Bronze & -- & -- & -- & -- & \textbf{0.043*} & \cellcolor{small}0.034 \\
\hline
% Gunning Fog section
\multirow{3}{*}{Gunning Fog}
& Gold & -- & -- & \textbf{0.017*} & \cellcolor{medium}0.124 & \textbf{0.032*} & \cellcolor{medium}0.088 \\
& Silver & -- & -- & -- & -- & \textbf{0.030*} & \cellcolor{medium}0.107 \\
& Bronze & -- & -- & \textbf{0.024*} & \cellcolor{small}0.046 & \textbf{0.049*} & \cellcolor{small}0.031 \\
\hline
% Code Smells section
\multirow{3}{*}{Code Smells}
& Gold & -- & -- & -- & -- & -- & -- \\
& Silver & -- & -- & -- & -- & \textbf{0.004**} & \cellcolor{large}0.209 \\
& Bronze & -- & -- & -- & -- & -- & -- \\
\hline
% Technical Debts section
\multirow{3}{*}{Technical Debts}
& Gold & -- & -- & -- & -- & -- & -- \\
& Silver & -- & -- & -- & -- & \textbf{0.004**} & \cellcolor{large}0.209 \\
& Bronze & -- & -- & -- & -- & -- & -- \\
\hline
% Violations section
\multirow{3}{*}{Violations}
& Gold & -- & -- & -- & -- & -- & -- \\
& Silver & -- & -- & -- & -- & \textbf{0.006**} & \cellcolor{large}0.191 \\
& Bronze & -- & -- & -- & -- & -- & -- \\
\hline
\end{tabular}
\end{table*}
% Required packages:
% \usepackage{booktabs}
% \usepackage[table]{xcolor}
% \usepackage{graphicx}

% Define colors for alternating rows
\definecolor{lightblue}{RGB}{240,248,255}
\definecolor{mediumblue}{RGB}{173,216,230}
\begin{table*}[htbp]
% \vspace*{-1.62em}
\centering
\caption{Human vs GenAI Feature Statistics Comparison in IEEE-CIS Competition}
\label{tab:ai-human-stats}
\small
\renewcommand{\arraystretch}{1.3}
\setlength{\tabcolsep}{16pt}

\begin{tabular}{llcccc}
\hline
\textbf{Feature} & \textbf{Source} & \textbf{Mean} & \textbf{Median} & \textbf{Min} & \textbf{Max} \\
\hline

\multirow{2}{*}{Functions} & Human & 3.16 & 2.0 & 0 & 46 \\

 & \cellcolor{lightblue}AI & \cellcolor{lightblue}0.67 & \cellcolor{lightblue}0.0 & \cellcolor{lightblue}0 & \cellcolor{lightblue}2 \\
\hline

\multirow{2}{*}{Statements} & Human & 121.49 & 91.5 & 10 & 586 \\

 & \cellcolor{lightblue}AI & \cellcolor{lightblue}94.33 & \cellcolor{lightblue}35.0 & \cellcolor{lightblue}25 & \cellcolor{lightblue}223 \\
\hline

\multirow{2}{*}{Comment Lines} & Human & 40.93 & 23.5 & 0 & 252 \\

 & \cellcolor{lightblue}AI & \cellcolor{lightblue}32.67 & \cellcolor{lightblue}18.0 & \cellcolor{lightblue}13 & \cellcolor{lightblue}67 \\
\hline

\multirow{2}{*}{\#Code Cell} & Human & 31.36 & 22.0 & 2 & 205 \\

 & \cellcolor{lightblue}AI & \cellcolor{lightblue}9.00 & \cellcolor{lightblue}8.0 & \cellcolor{lightblue}7 & \cellcolor{lightblue}12 \\
\hline

\multirow{2}{*}{Cyclomatic Complexity} & Human & 14.30 & 9.5 & 0 & 126 \\

 & \cellcolor{lightblue}AI & \cellcolor{lightblue}14.67 & \cellcolor{lightblue}0.0 & \cellcolor{lightblue}0 & \cellcolor{lightblue}44 \\
\hline

\multirow{2}{*}{Gunning Fog} & Human & \textcolor{red}{\textbf{10.32}} & 9.91 & 0 & 64.61 \\

 & \cellcolor{lightblue}AI & \cellcolor{lightblue}16.07 & \cellcolor{lightblue}16.13 & \cellcolor{lightblue}11.55 & \cellcolor{lightblue}20.53 \\
\hline

\multirow{2}{*}{Code Smells} & Human & 29.25 & 0.0 & 0 & 3166 \\

 & \cellcolor{lightblue}AI & \cellcolor{lightblue}\textcolor{blue}{\textbf{3.00}} & \cellcolor{lightblue}3.0 & \cellcolor{lightblue}3 & \cellcolor{lightblue}3 \\
\hline

\multirow{2}{*}{Technical Debts} & Human & 44.39 & 0.0 & 0 & 3228 \\

 & \cellcolor{lightblue}AI & \cellcolor{lightblue}\textcolor{blue}{\textbf{15.00}} & \cellcolor{lightblue}15.0 & \cellcolor{lightblue}15 & \cellcolor{lightblue}15 \\
\hline

\multirow{2}{*}{Violations} & Human & 29.59 & 0.0 & 0 & 3167 \\

 & \cellcolor{lightblue}AI & \cellcolor{lightblue}\textcolor{blue}{\textbf{3.00}} & \cellcolor{lightblue}3.0 & \cellcolor{lightblue}3 & \cellcolor{lightblue}3 \\
\hline
\end{tabular}
\begin{minipage}{\linewidth}
\centering
\vspace*{1em}
\footnotesize
Color coding indicates superior performance: \textcolor{blue}{\textbf{AI outperforms}} vs. \textcolor{red}{\textbf{Human outperforms}}.
\end{minipage}
\end{table*}
% \vspace{-0.5em}

To investigate distinguishing characteristics between GenAI and human-written data science notebooks, we conducted statistical analyses using the Kruskal-Wallis H test \cite{kruskal1952use} across the human-written notebooks dataset and the GenAI notebook dataset, as illustrated in Figure \ref{fig:overview}, focusing on medium ($\geq$0.06) to large ($\geq$0.14) effect sizes that indicate practically meaningful differences.

\textbf{Documentation Accessibility}: Table~\ref{tab:ai-human-stats} reveals that human-written notebooks consistently produced more accessible documentation despite GenAI's code quality advantages. The statistical comparison shows GenAI explanatory text in the IEEE-CIS competition consistently operated at senior-college-level complexity (Gunning Fog score: 16.07), compared to human documentation which operated at high-school-senior level (Gunning Fog score: 10.32). Statistical analysis in Table \ref{tab:ai-each} confirmed these readability differences across multiple competitions, with medium effect sizes observed for both gold and silver medal comparisons in IEEE-CIS (p-value $\leq$ 0.05, $\eta^{2}$ $\geq$ 0.088), while Santander showed similar patterns (p-value = 0.017, $\eta^{2}$ = 0.124). Furthermore, GenAI notebooks showed significantly fewer comment lines compared to human-written notebooks in the Home-Credit competition (p-value = 0.035, $\eta^{2}$ = 0.122, medium effect size), indicating reduced inline documentation coverage alongside the increased complexity of explanatory text.

\textbf{Code Quality Features}:  Conversely, Table~\ref{tab:ai-each} shows that GenAI notebooks demonstrated superior performance in technical code quality measures. In the IEEE-CIS competition, GenAI notebooks significantly outperformed human-written notebooks across several code quality indicators with large effect sizes ($\geq$ 0.191): code smells (p-value = 0.004, $\eta^{2}$ = 0.209), technical debt (p-value = 0.004, $\eta^{2}$ = 0.209), and violations (p-value = 0.006, $\eta^{2}$  = 0.191). Similarly, the Home-Credit competition showed GenAI notebooks exhibiting reduced structural complexity compared to medal-winning submissions, with large effect sizes observed for functions (GenAI mean: 4.00 vs Human mean: 5.71, p-value = 0.011, $\eta^{2}$ = 0.197) and statements (GenAI mean: 144.33 vs Human mean: 194.38, p-value = 0.025, $\eta^{2}$ = 0.143).Additionally, medium effect sizes were found for the number of code cells (GenAI mean: 13.00 vs Human mean: 33.65, p-value = 0.049, $\eta^{2}$ = 0.103) and cyclomatic complexity (GenAI mean: 16.00 vs Human mean: 18.53, p-value $\leq$ 0.050, $\eta^{2}$ $\geq$ 0.102), indicating GenAI generates more streamlined code organization with lower algorithmic complexity.

\textbf{Execution and Reproducibility}: 
Of the nine notebooks executed, there were three that failed to complete the execution. These execution outcomes confirm that static analysis alone is insufficient, as it misses critical data-handling and logic errors that limit the notebooks' practical utility.  The llama\_home-credit and llama\_IEEE notebooks failed due to runtime errors from passing non-numeric columns to the model. The GPT\_IEEE notebook failed because column names were not consistently unified, leading to key mismatches during data processing. 

The analysis demonstrates that GenAI and human notebooks exhibit distinct but complementary strengths: GenAI excels in producing clean, standardized code with superior adherence to best practices, while humans demonstrate greater structural innovation and complexity. Despite GenAI's advantages in code quality metrics, including lower technical debt, fewer code smells, and reduced violations, medal-worthy performance appears to depend more on human strengths such as creative problem-solving, domain expertise, and algorithmic innovation rather than code cleanliness alone.

\begin{tcolorbox}[colback=gray!5,colframe=awesome,title= RQ2 Summary]
Human and GenAI notebooks have statistical differences. 
We identify and summarize key differences below:
\begin{itemize}
    \item \textit{Documentation features -} GenAI notebooks are harder to read (higher reading levels), while the human written documentation is easier to read (gunning fog)
    \item \textit{Code features -} GenAI notebooks demonstrate higher code quality with fewer code smells, lower technical debt, and reduced violations. Meanwhile, human-written code is more complex, with greater structural complexity and higher function counts and statements.
\end{itemize}
\end{tcolorbox}

\section{Discussion}

In this section, we present a discussion of the implications and the research agenda from the three case studies.

\subsection{Implications}

This study provides a comparative analysis of human-written and the raw output of GenAI without human intervention data science notebooks in competitive environments for an understanding of the baseline capabilities and inherent strengths and weaknesses of both approaches. 
Our findings reveal distinct complementary strengths between Human and GenAI approaches that suggest strategic applications in data science practice. 
As a proof of concept, there are differences, and we believe that a more comprehensive study will highlight more differences between the Human and AI. 
Although preliminary, the result from RQ2 demonstrates that GenAI notebooks achieved significantly superior code quality metrics features (code smells, technical debt, and violations). These results indicate that GenAI tools could serve as valuable automated code review and standardization tools.

The results from RQ2 also reveal a limitation: GenAI documentation operated at senior-college reading levels (Gunning Fog: 16.07) compared to human high-school-senior levels (10.32). Combined with our RQ1 findings showing that documentation features had the strongest discriminating power between gold and non-gold medalists, this suggests that current GenAI limitations in accessible documentation may undermine competitive performance despite superior code quality.
In this study, we use the gold medal, but other heuristics of quality could be employed in future studies. 
Hybrid approaches, where GenAI assists in routine code generation and quality assurance while humans focus on documentation and complex problem-solving, may offer the best results. For immediate research directions, we will need to experiment with a larger and more diverse dataset that needs to be collected.
In this study, we only provided 3 competitions, so we need more competitions. 
We would also need to interview developer and prototype different tools that simulate the human to GenAI collaboration.
Would a competitor or the competition host be able to tell the difference between a human and GenAI solution.

The implications from the study at this stage are threefold. The first is that for hosts of the competitions, there are distinguishable features of GenAI usage, thus could be used to detect when GenAI could be used when not permitted. For competitors, these results provide insights into how to best use the GenAI-assistants in their toolbox. 
For researchers, the results may imply what are the skills necessary and cannot be amplified by use of GenAI.

This study provides a foundation for further exploration of the intersections between GenAI-assisted development, human creativity, and competitive data science performance. RQ1 and RQ2 highlight the need to balance code quality with documentation comprehensiveness in evaluating excellence. To build on these insights, we develop the following agenda as a roadmap:
\\
\begin{itemize}
    \item \textbf{Agenda One. Explore notebooks as an IDE for Human-GenAI documentation and code collaboration} Investigate whether computational notebooks are truly the optimal environment for human–GenAI collaboration, or if alternative development contexts might better support creative and effective hybrid work.
    \item \textbf{Agenda Two. Conduct a comprehensive study} Expand analysis to larger and more diverse datasets, encompassing different competitions and platforms beyond Kaggle, to validate whether findings such as the centrality of documentation quality hold across domains.
    \item \textbf{Agenda Three. Expand from Notebooks to Software Projects} Extend the study to traditional software artifacts, including source code repositories and developer documentation, to determine whether the trade-offs between code quality and documentation clarity persist in broader software engineering settings. Additional questions that could arise as secondary research topics could explore how different AI documentation is from human documentation in code comments, documentation artifacts and in discussions with other team members. 
    \item \textbf{Agenda Four. Explore challenges and risks of using GenAI} Examine the ethical and procedural challenges introduced by GenAI, particularly the risks of GenAI-augmented cheating and violations of competition rules. Future research should develop guidelines and detection mechanisms to ensure fairness and maintain trust in both competitive and collaborative environments.
\end{itemize}

\section{Threats to Validity}
This section examines potential threats to the validity and measures implemented.

\subsection{Internal Validity}
Three primary threats to internal validity require consideration.
\begin{itemize}
\item First, selection bias affects both RQ1 and RQ2 as our dataset comprises only medal-winning notebooks from three Kaggle competitions (Santander Customer Transaction Prediction, Home Credit Default Risk,
and IEEE-CIS Fraud Detection), which may not represent broader data science practices. This could bias our Kruskal-Wallis H test results comparing gold vs non-gold medalists and our human vs GenAI statistical comparisons. We address this by analyzing 465 notebooks across multiple competition domains.

\item Second, the potential contamination of our human-written dataset with AI-generated content in both RQ1 and RQ2, as notebooks for older competitions may have been created recently. While the purity of this dataset cannot be guaranteed. However, this concern is mitigated by the large, observed gap in code quality between the human and GenAI groups, which suggests the comparison remains meaningful.

\item Three, confounding variables like participant experience may simultaneously influence documentation practices and competition success in RQ1, while in RQ2, human notebooks represent experienced medal winners compared against GenAI lacking domain expertise. This could potentially confound our comparisons of code quality metrics and readability measures like Gunning Fog scores. The use of objective, quantifiable metrics helps to mitigate this concern.
\end{itemize}

\subsection{External Validity}
Three threats to external validity merit discussion.
\begin{itemize}
\item Generalizability beyond Kaggle competitions may be limited. Our RQ1 findings that documentation features are the strongest discriminators between gold and non-gold medalists and RQ2 results showing human superiority in documentation accessibility versus GenAI superiority in code quality may not apply to industry contexts where performance metrics differ. However, these patterns likely extend beyond competitions as they reflect fundamental differences in how current GenAI systems generate code versus explanatory text.
\item Temporal validity affects our GenAI methodology. The models used in RQ2 represent current capabilities but could be outdated in the future. We mitigate this by focusing on fundamental trade-offs between code standardization and human-readable explanation revealed across all three competitions.
\item We acknowledge that the sample size of 9 GenAI-generated notebooks constitutes a limitation of this study. This number was determined by the exploratory nature of the research and the computational resources required for generation and analysis. While the findings provide initial insights, a larger and more diverse sample of GenAI notebooks is necessary to ensure the generalizability of the results. Future work should aim to expand this dataset significantly.
\end{itemize}

\subsection{Construct Validity}
One threat to construct validity requires acknowledgment.
\begin{itemize}
\item Medal-worthy performance definition: Our first construct, notebook quality, is proxied by Kaggle's medal system (i.e., Bronze, Silver, Gold). This is a potential threat, as competition rankings may not capture all dimensions of exceptional data science work, such as long-term maintainability, real-world business impact, or deployment efficiency outside the competition environment. However, we mitigate this by using the established Kaggle medal criteria, which represent community-validated standards for excellence in competitive data science. This provides a reliable and holistic benchmark for notebook quality within the competitive data science context.
\item Practical Usability of Code Quality Metrics: Our initial analysis defined code quality using static metrics like code smells and technical debt, which was a valid concern regarding construct validity as it does not capture runnability. To address this and clearly bound our claims, we conducted a post-review execution validation. As reported in Section IV-B, this check confirmed that three of GenAI notebooks failed to run, despite their high static quality. This finding reinforces the limitation and shows that our 'code quality' construct does not fully map to 'practical usability'.
\end{itemize}

\section{\textbf{Related Work}}

\subsection{Studies on Computational Notebooks}

Prior research has established computational notebooks as a principal paradigm for data scientists, valued for their support of two distinct roles: private exploration and public explanation \cite{10.1145/3173574.3173606}. As literate programming environments, they are designed to integrate executable code, narrative documentation, and visualizations within a single document \cite{10.1145/3173574.3173748, 10.1145/3313831.3376729}. However, achieving narrative coherence is not an automatic outcome. Kery et al. \cite{10.1145/3173574.3173748} underscore the significant human effort involved, observing that data scientists find it a non-trivial task to clean up and curate their messy, exploratory code into coherent ``stories'' for presentation or sharing .

In collaborative settings, a well-constructed narrative is indispensable for communicating results and knowledge dissemination. This dynamic is particularly evident in competitive platforms like Kaggle. Wang et al. \cite{10.1145/3411763.3451617} demonstrated that community-defined quality (i.e., highly-voted notebooks) correlates more strongly with extensive documentation and high readability than with the notebook's objective performance score on the leaderboard . Their analysis of these well-documented notebooks revealed nine distinct categories of documentation, including process descriptions, headlines for navigation, and explanations for analytical reasoning . This finding underscores the role of notebooks as vital communication artifacts, not merely as code containers.

Yet, this ideal of a clean, well-documented notebook is often at odds with the reality of data science workflows. The transition from exploration to explanation is fraught with pain points. Chattopadhyay et al. \cite{10.1145/3313831.3376729} cataloged numerous challenges that disrupt data scientists, including difficulties in setting up environments, managing dependencies, versioning exploratory code, and deploying notebooks to production. This difficulty contributes to the well-documented notebook-to-production gap \cite{9793798}.

A critical aspect of this gap is the widespread lack of reproducibility. A large-scale study of notebooks from biomedical publications by Samuel et al. found that the "large majority" of notebooks could not be automatically executed. The primary cause was "issues with the documentation of dependencies" , with most notebooks failing due to \textbf{ModuleNotFoundError} or \textbf{ImportError} . That same study noted a key correlation: notebooks with a low ratio of markdown-to-code cells were more likely to have exceptions, directly linking a lack of documentation to poor reproducibility . This challenge is precisely what Quaranta et al. \cite{9793798} identified as the key blocker: deficiencies in code quality, structure, and maintainability prevent notebooks from being reliably integrated into production workflows.

\subsection{Human and GenAI Complementarity}

This emphasis on human-driven documentation, narrative, and high-level strategy aligns with findings on human-GenAI complementarity. Research indicates that humans demonstrate superior performance in areas requiring deep domain expertise, complex logical reasoning, and creative solutions. For example, Licorish et al. \cite{licorish2025comparinghumanllmgenerated} found that humans performed better on tasks requiring in-depth domain knowledge, such as quantum optimization algorithms or debugging complex logic. GenAI, in contrast, specializes in the rapid generation of structured content and the management of repetitive, boilerplate code~\cite{bangerl2024explorations}. How this complementary relationship manifests specifically within computational notebooks remains underexplored.

The introduction of GenAI is prompting a re-conceptualization of the developer's role. Rather than authoring code line-by-line, the human is repositioned as a ``curator'' \cite{cotroneo2025humanwrittenvsaigeneratedcode} or as a ``system orchestrator'' responsible for providing high-level intent ~\cite{treude2025generative,marron2024new,hassan2024towards}. In this paradigm, the human directs the overall strategy and validates the AI's output, while the AI manages the low-level implementation. Consequently, human validation emerges as a critical function.

Our research builds upon this concept by quantitatively comparing how this role differentiation manifests in a competitive environment. We examine the balance between documentation (a human strength) and technical code quality (a GenAI strength). Our work is supported by findings from Molison et al. \cite{molison2025llmgeneratedcodemaintainable}, which found that code generated by LLMs generally contains fewer bugs and requires less remediation effort. They also observed that fine-tuning, while effective at reducing high-severity blocker and critical bugs by shifting them to lower-severity categories, simultaneously degraded the model's overall performance. Moreover, for complex, competition-level tasks, LLMs were found to introduce structural problems not present in human-authored code. Similarly, Licorish et al. \cite{licorish2025comparinghumanllmgenerated} found that GPT-4 passed a higher percentage of functional test cases than human-written code. Cotroneo et al. \cite{cotroneo2025humanwrittenvsaigeneratedcode} add to this, noting that AI code is generally simpler and more repetitive.

% \section{Conclusion}
% This study conducted a comparative analysis of human-written and GenAI notebooks within the context of competitive data science to explore the distinct strengths of each. The findings reveal a clear and complementary set of advantages. GenAI excels at producing code with high static technical quality, featuring fewer code smells and less technical debt. However, our execution checks revealed this does not always translate to practical runnability, with several notebooks failing due to data-handling errors. In contrast, human notebooks, particularly medal-winning entries, are distinguished by their comprehensive and accessible documentation.

% The results suggest that while GenAI can achieve superior code cleanliness, top-tier performance in these competitions currently depends more on human-centric strengths like creative problem-solving, algorithmic innovation, and insightful documentation rather than on code quality alone. The implications of these findings are threefold: they may allow competition hosts to identify features of GenAI usage, guide competitors in how to best use GenAI assistants, and help researchers understand which skills are not easily amplified by current AI tools. Ultimately, this work provides a foundation for exploring the intersections of GenAI development and human creativity, pointing toward a future of hybrid approaches where GenAI handles routine code generation while humans drive the innovative problem-solving and narrative communication that define exceptional data science.

\section{Conclusion}
This study conducted a comparative analysis of human-written and GenAI-generated notebooks within the high-stakes context of competitive data science. Our goal was to explore the distinct, and often complementary, strengths and weaknesses inherent in both human and AI approaches. The findings reveal a clear dichotomy: GenAI excels at producing code with high static technical quality, consistently featuring significantly fewer code smells and less technical debt than human-written counterparts. This suggests a strong capability for adhering to programming standards and generating clean, standardized code.

However, this technical superiority comes with a critical caveat. Our execution checks revealed that high static quality does not translate to practical runnability, as several GenAI notebooks failed during execution due to fundamental data-handling and logic errors. In contrast, human-written notebooks, particularly the medal-winning entries, were distinguished by their comprehensive, accessible, and insightful documentation. This aligns with our RQ1 findings, confirming that narrative communication and clear explanations are a hallmark of high-quality human work in this domain.

The results strongly suggest that while GenAI can achieve superior code cleanliness, top-tier performance in these competitions currently depends more on human-centric strengths. These include creative problem-solving, domain-specific algorithmic innovation, and the ability to craft insightful documentation that communicates a clear analytical narrative—skills that code quality metrics alone fail to capture.

The implications of these findings are threefold and significant for the data science community:
\begin{itemize}
    \item \textbf{For competition hosts:} The identifiable features of GenAI-generated code (e.g., high static quality but low readability scores) may allow for the development of new heuristics to detect GenAI usage, which is crucial for upholding competition integrity and rules.
    \item \textbf{For competitors:} These results provide a strategic guide for using GenAI assistants. Competitors can leverage these tools to handle routine tasks like code standardization and boilerplate generation, freeing up their own time to focus on high-impact, human-centric tasks such as feature engineering, model innovation, and narrative-building.
    \item \textbf{For researchers and educators:} Our findings help identify which critical data science skills, such as narrative reasoning, complex problem decomposition, and practical validation, are not easily amplified or replicated by current AI tools, guiding future research and curriculum development.
\end{itemize}

Ultimately, this work provides a foundation for exploring the intersections of GenAI development and human creativity. It points toward a future of hybrid approaches where GenAI handles routine code generation and quality assurance, while humans drive the innovative problem-solving and narrative communication that define exceptional data science.

\section{Acknowledgement}
This research is supported by JSPS Kakenhi (A) JP24H00692.

\section{Data Availability}
\label{section:data_availability}
To facilitate replication studies, our scripts and dataset are publicly available online, which can be found at \DOIbox{10.5281/zenodo.15606124}

\bibliographystyle{IEEEtranS.bst}
\bibliography{reference}

\end{document}